# Exploratory analysis of a measurement scale of an information security management system

Rúsbel Domínguez Domínguez[1], Omar Flores Laguna[2], José A. Sánchez-Valdez[3]

[12] Facultad de Ciencias Empresariales y Jurídicas, Universidad de Montemorelos, Montemorelos, México.
[3] Logística, Calidad y Manufactura, Tecnológico Nacional de México, Monterrey, México.

rusbel@um.edu.mx, oflores@um.edu.mx, jasanchez0777@gmail.com

## 1    Introduction

Currently, information and data is the most valuable asset for institutions. This progress in computing and communication networks, requires a higher level of prevention and responsibility in the face of cyber threats [1, 2].

Information security has to do with risk management since ISMS standards allow measuring threats and vulnerable targets of information systems in organizations, which allows taking actions against such threats. This is why if an institution fails in information security management, the integrity of its data will be compromised and its finances could be affected [3, 4].

For this reason, the cybersecurity scenario has forced organizations to incorporate a set of good security practices in their information management systems.

These protection practices have led different organizations to define and implement information security standards [4, 5].

Risk management is an important element of the strategic management of institutions and, on many occasions they are crucial to systematize business activities and continue operations. It is necessary to point out the guidelines to properly manage the risk of organizations [6, 7].

In a study conducted at an Asian university, the probability of threats and damage to the confidentiality, integrity and availability of information has never been higher. Thanks to this project, they became aware of the information security of their assets (IT infrastructure, records, research data and student information), adopting information security best management practices (ISMS) based on the ISO/IEC 27001:2013 standard [8].

On the other hand, [9] conducted an ISMS model for basic level educational institutions. They analyzed the critical assets of the academic secretariat area of educational institutions, based on the ISO/IEC 27001:2013 standard. The resulting model complied with the mandatory requirements, established by the standard that help ensure the availability, integrity and confidentiality of information.

The creation and validation of instruments that measure the degree of management in an Information Security Management System that allows to maintain systematized and standardized based on ISO/IEC 27002:2013 standards, help the analysis and evaluation of risks in the IT assets of an organization and through which the availability, integrity and reliability in the management of a company's information is guaranteed. This scale has 24 items, divided into four factors, policies and regulations of the organization, privacy, integrity and authenticity.

This research is organized as follows: (a) theoretical framework, (b) methodology, (c) results and (d) discussion and conclusions.

## 2    Related work

ISO 27001 is a standard program that allows the implementation of security mechanisms to protect information systems as important assets of organizations. The ISO 27001 describes the guidelines that organizations must have to ensure the availability, integrity and authenticity as well as the confidentiality of information [10].

### 2.1. Information Security Management System (ISMS)

What is the information security management system? The ISMS is a systematized and standardized process based on the ISO/IEC 27002:2013 standards, which allows analyzing and assessing the risks of an organization's IT assets and through which the availability, integrity and reliability in the management of a company's information is guaranteed [11].

An information security management system (ISMS) can be defined as a management system used to maintain and establish a secure information environment. The ISMS considers the maintenance of processes and procedures to manage information technology security. These actions contemplate the need to identify information security vulnerabilities as well as implement strategies that help minimize risks, know the needs, measure the results and improve protection strategies [12].

An ISMS consists of a set of policies, instructions and procedures specific to each information system that aims to protect information assets in an institution [13].
Procedures are usually characterized by having several activities executed in a certain order and require a chain of resources (equipment, facilities and personnel), as well as a series of inputs to obtain a final result or output. This is what we call "processes in an organization" [14].

## 3    Dimensions

For an information security management system most authors [11, 15–26] agree that the dimensions (domains) of ISO 27002:2013 since the 2005:2013 revision are classified into eleven dimensions: (a) security policies, (b) security organization, (c) asset management, (d) human resource security, (e) physical and environmental security, (f) communications and operations management, (g) access control, (h) information systems acquisition, development and maintenance, (i) security incident management, (j) business continuity management and (k) compliance.

Other authors [10, 27–30] agree that the dimensions (domains) of ISO 27002:2013 since the 2005:2013 revision are classified into 14 dimensions: (a) Security Policies, (b) Information Security Organization, (c) Human Resource Security, (d) Asset Management, (e) Access Control, (f) Encryption, (g) Physical and Environmental Security, (h) Operations Security, (i) Telecommunications Security, (j) Information Systems Acquisition, Development and Maintenance, (k) Supplier Relationship, (l) Information Security Incident Management, (m) Information Security Aspects of Business Continuity Management, (n) Compliance.

[10] mention four dimensions (domains): (a) policies and regulations, (b) privacy, (c) integrity and (d) authenticity. For this research, the categorization proposed by the aforementioned authors was used.

### 3.1. Proposed reagents

Presents the items of the IM-SGSI scale (Table 1)

**Table 1** IM-SGSI factors and items

| Items | Authors |
|---|---|
| **Organizational policies and regulations** | |
| 1. Information security policy manual. | [10, 11, 18–20, 23, 27, 28, 30] |
| 2. Periodic reviews of information security policies. | [11, 15, 17–20, 22, 28, 31, 32] |
| 3. Assignment of responsibilities for information security | [10, 11, 15, 17–20, 23, 27, 30] |
| 4. Policies for the use of wireless networks | [11, 18, 19, 32] |
| 5. The implementation of terms and conditions in the labor contract for personnel. | [11, 15, 18, 20, 27, 30, 32] |
| 6. Information Security Awareness, Education and Training | [10, 11, 18–20, 22, 27, 30] |
| 7. Planning for business continuity during adverse events | [10, 11, 15, 17–20, 22, 23, 27, 28, 30–32] |
| 8. Compliance with legal and contractual requirements | [11, 15, 18, 20, 27, 30, 32] |
| 9. Periodic reviews of information security | [10, 11, 15, 17–20, 23, 27, 30–32] |
| **Privacy** | |
| 10. Responsible use of network equipment and servers | [10, 15, 19, 20, 23, 31, 32] |
| 11. Backing up and encrypting files | [10, 19, 20, 22, 23, 29, 32] |
| 12. An access control policy for the facilities, nodes and data center | [10, 11, 15, 17–20, 22, 23, 27, 29–32] |

*(continued)*

**Table 1** (continued)

| Items | Authors |
|---|---|
| 13. System administration tools (SolarWinds, Active Directory, Nagios, etc.) | [10, 11, 19, 20, 22, 23, 27, 28, 32] |
| 14. Restrictions on access to information | [10, 11, 15, 17–20, 22, 23, 27, 29–32] |
| **Integrity** | |
| 15. A policy on the use, protection and lifecycle of access keys | [10, 11, 19, 22, 23, 27, 30] |
| 16. Protection against external and environmental threats | [10, 11, 15, 17–20, 22, 23, 27, 32] |
| 17. Documentation of operating procedures | [10, 11, 15, 17, 20, 27–32] |
| 18. Division of networks according to groups of services, users and information systems | [10, 11, 15, 19, 20, 23, 27–30, 32] |
| 19. A policy for the development and acquisition of secure software | [10, 11, 15, 17–20, 22, 23, 27–29, 32] |
| 20. An oversight and review of services provided by third parties | [10, 11, 15, 18–20, 23, 30, 32] |
| **Authenticity** | |
| 21. Effective and consistent management of information security incidents. | [10, 11, 15, 17–20, 22, 23, 28–32] |
| 22. Notifications of information security events | [10, 15, 17–20, 22, 23, 27, 28, 30–32] |
| 23. A response to security incidents | [10, 11, 15, 17–20, 22, 23, 28–32] |
| 24. A collection of evidence of security incidents | [10, 11, 15, 18–20, 27, 30–32] |

# 4    Methodology

The population used in this research consisted of students studying at two universities in northeastern Mexico. Those evaluated were subjects of legal age, of indistinct gender, studying the branches of engineering (systems, industrial and systems, information and communication technologies, electronics and telecommunications). The sample excluded people who are not studying engineering, graduates and consequently, people who do not belong to either of the two institutions where this instrument was applied.

Due to the health environment currently being experienced in the world due to the pandemic (COVID-19), the instrument was applied through the "Google Forms" platform. It was not necessary to ask participants for personal information, nor to register their name or e-mail address, in order to respect their identity and maintain confidentiality. Subjects who did not wish to participate in the research simply ignored the form. The participants of this instrument had access to informed consent before answering the survey. Access to the information provided by the participants is completely confidential. The sample consisted of 143 participants, of which 42 were students from the Universidad de Montemorelos and 101 students from the Tecnológico de Nuevo León.

### 4.1. The instrument

The Information Security Management System (ISMS) scale was developed in-house and has 24 items divided into four factors:

organizational policies and regulations (PR1 to PR9), privacy (P10 to P14), integrity (I15 to I20) and authenticity (A21 to A24). A criterion followed in this research is that mentioned by [33] the factors should have a minimum of 3 or 4 items per factor and a minimum of 200 cases.

## 5    Analysis of results

To determine construct validity, an exploratory factor analysis (EFA) was carried out using Jamovi software version 1.2.27.

### 5.1. Descriptive statistics

Within the descriptive statistics, skewness and kurtosis were calculated for each item, in addition, the Shapiro Wilks test was obtained for the items used (Table 2). As can be seen, most of the data for each item do not meet the range criterion (-1 to 1) for univariate normality of the items [34].

### 5.2. Exploratory factor analysis

Within Exploratory factor analysis was performed since the model is considered to be reflective, meaning that the items are the independent variables [35]. Similar instrument validation studies have used the principal components method, but this method should be used in formative models, that is, when the items or variables are continuous and independent [36].

The unweighted least squares (ULS) method was used for this study, which is the equivalent method of least residual (Minimun Residual) estimation [37].

[38] mention several authors [33, 39–43] who recommend using oblique rotation instead of varimax rotation.

Within the oblique rotation, there are the direct oblimin and promax methods, for this research it was decided to do a promax rotation.

To determine construct validity, factor analysis was used (KMO = .950 Bartlett's Sphericity significant < .001).

Table 3 presents information comparing the relative saturations of each item for the four IM-SGSI factors.

**Table 2**  Item asymmetry and kurtosis

| Items | *Asymmetry* | Kurtosis | Shapiro Wilks (p) |
|---|---|---|---|
| Information Security Policy Manual (PR1) | -0.46 | -0.69 | <.001 |
| Periodic reviews of the information security policies (PR2) | -0.38 | -0.51 | <.001 |
| Assignment of responsibilities for information security (PR3) | -0.51 | -0.66 | <.001 |
| Policies for the use of wireless networks (PR4) | -0.58 | -0.71 | <.001 |
| Implementation of terms and conditions in the labor contract for personnel (PR5) | -0.51 | -0.80 | <.001 |
| Information security awareness, education and training (PR6) | -0.47 | -0.62 | <.001 |
| Planning for business continuity during adverse situations (PR7) | -0.45 | -0.56 | <.001 |
| Compliance with legal and contractual requirements (PR8) | -0.54 | -0.65 | <.001 |
| Periodic information security reviews (PR9) | -0.40 | -0.91 | <.001 |
| Responsible use of network and server equipment (P10) | -0.70 | -0.35 | <.001 |
| Backing up and encrypting files (P11) | -0.39 | -0.70 | <.001 |
| A policy for access control to facilities, nodes and data center (P12) | -0.56 | -0.74 | <.001 |
| System Administration Tools (SolarWinds, Active Directory, Nagios, etc.) (P13) | -0.31 | -0.98 | <.001 |
| Restrictions on access to information (Q14) | -0.59 | -0.67 | <.001 |
| A policy on the use, protection and lifecycle of access keys (I15) | -0.70 | -0.43 | <.001 |
| Protection against external and environmental threats (I16) | -0.49 | -0.67 | <.001 |
| Documentation of operating procedures (I17) | -0.45 | -0.73 | <.001 |
| Division of networks according to groups of services, users and information systems (I18) | -0.72 | -0.44 | <.001 |
| A policy for the development and acquisition of secure software (I19) | -0.48 | -0.74 | <.001 |
| A monitoring and review of services provided by third parties (I20) | -0.40 | -0.64 | <.001 |
| Effective and consistent management of information security incidents (A21) | -0.51 | -0.62 | <.001 |
| Notifications of information security events (A22) | -0.28 | -1.05 | <.001 |
| A response to security incidents (A23) | -0.48 | -0.79 | <.001 |
| A compilation of evidence of security incidents (A24) | -0.31 | -0.83 | <.001 |

The second factor (column 2 of Table 3) initially consisted of the following items: "Periodic reviews of information security policies" (PR2), "Information security policy manual" (PR1), "Assignment of responsibilities for information security" (PR3), "Implementation of terms and conditions in the labor contract for personnel" (PR5), "Awareness, education and training in information security" (PR6).

**Table 3** Factor loadings by oblique rotation with the promax method

| Items | Factor 1 | Factor 2 | Factor 3 | Factor 4 | Unique-ness |
|---|---|---|---|---|---|
| Planning for business continuity during adverse situations (PR7) | 0.919 | | | | 0.1474 |
| A response to security incidents (A23) | 0.828 | | | | 0.1988 |
| Compliance with legal and contractual requirements (PR8) | 0.775 | | | | 0.2222 |
| A compilation of evidence of security incidents (A24) | 0.726 | | | | 0.2022 |
| A monitoring and review of services provided by third parties (I20) | 0.692 | | | | 0.2449 |
| Information security policies and regulations in place (PR9) | 0.648 | | | 0.301 | 0.1513 |
| There is effective and consistent management of information security incidents (A21) | 0.555 | | | | 0.1705 |
| Notifications of information security events (A22) | 0.531 | | | 0.319 | 0.2567 |
| A policy for the development and acquisition of secure software (I19) | 0.464 | | | 0.406 | 0.2308 |
| Periodic reviews of the information security policies (PR2) | | 0.964 | | | 0.0956 |
| Information Security Policy Manual (PR1) | | 0.940 | | | 0.1634 |
| Assignment of responsibilities for information security (PR3) | | 0.733 | | | 0.2485 |
| Implementation of terms and conditions in the labor contract for personnel (PR5) | | 0.593 | | | 0.3564 |
| Information security awareness, education and training (PR6) | 0.399 | 0.427 | | | 0.3003 |
| A policy on the use, protection and life cycle of access keys (I15) | | 0.369 | | 0.358 | 0.3089 |
| A policy for access control to facilities, nodes and data center (P12) | | | 0.723 | | 0.2197 |
| Policies for the use of wireless networks (PR4) | | | 0.698 | | 0.3848 |
| Backing up and encrypting files (P11) | | | 0.657 | | 0.2420 |
| Responsible use of network and server equipment (P10) | | | 0.626 | | 0.3030 |
| System Administration Tools (SolarWinds, Active Directory, Nagios, etc.) (P13) | 0.391 | | 0.511 | | 0.4203 |
| Restrictions on access to information (Q14) | | | | | 0.6095 |
| Division of networks according to groups of services, users and information systems (I18) | | | | 0.794 | 0.2136 |
| Documentation of operating procedures (I17) | | | | 0.727 | 0.1954 |
| Protection against external and environmental threats (I16) | | | | 0.530 | 0.2951 |

Items (PR7, PR8, PR9) were grouped into the "authenticity" factor, while item (PR4) was grouped into the "privacy" factor. It was decided to change the wording of these items and leave them in the factor initially proposed, which is "policies and regulations". The items were reworded as follows: "There are business continuity planning policies for business continuity during adverse situations" (PR7), "There are policies and regulations for compliance with legal and contractual requirements" (PR8), "There are information security policies and regulations" (PR9), "There are policies and regulations for the use of wireless networks" (PR4).

The third factor (column 3 of Table 3) grouped all the items of the "privacy" dimension (P10 to P14), the items grouped by their factor loadings were the following: "A policy of access control to facilities, nodes and data center" (P12), "The backup and encryption of files" (P11), "The responsible use of network and server equipment" (P10), "System administration tools (SolarWinds, Active Directory, Nagios, etc.)" (Q13), "Information access restrictions" (Q14). Although this item has a factor loading of less than .30, it was grouped in its corresponding factor.

The fourth factor (column 4 of Table 3) was constituted after the rotation with three of the six items, the items grouped by their factor loadings were the following: "Division of the networks according to groups of services, users and information systems" (I18), "Documentation of operating procedures" (I17), "Protection against external and environmental threats" (I16).

Although items (I19) and (I15) have a higher loading (very minimal) in other factors, it was decided to make a small adjustment in the wording and leave them in their corresponding factor, where they also have a very important factor loading, the items were worded as follows: "There is an effective evaluation for the development and acquisition of secure software" (I19), "There is a correct management, protection and life cycle of access keys" (I15). The item (I20) has an important factorial load in the "authenticity" factor, it was decided to reword it as follows: "There is an integrated supervision and review of the services provided by third parties" and leave it in its original dimension.

## 6 Composite reliability (Cronbach's Alpha and McDonalds' Alpha)

To calculate the reliability of the instrument, the composite reliability (CR) was used using the McDonald omega coefficient. The choice of this coefficient is based on different researchers [44, 45] who explain that this index should be used. According to [46], the omega coefficient works with the factor loadings and this makes the calculations more stable, reflecting the true level of reliability. [44] mentions that the omega coefficient is not affected by the number of items. To consider an acceptable reliability value using the omega coefficient, these should be between .70 and .90 [47].

When applying the composite reliability, the results of the omega coefficient for the factors were as follows: (a) policies and regulations (PR) was equal to. 947, (b) authenticity (A) was equal to. 932, (c) integrity (I) was .936 and (d) privacy (P) was equal to .892. Thus, it is shown that reliability is acceptable in all factors (Table 4).

**Table 4** Reliability scale

|  | Cronbach's α | McDonald's ω |
|---|---|---|
| Policies and regulations (PR) | 0.946 | 0.947 |
| Authenticity (A) | 0.931 | 0.932 |
| Integrity (I) | 0.936 | 0.936 |
| Policy (P) | 0.887 | 0.892 |

## 7    Discussion

The purpose of this research is to propose an instrument to measure the degree of an information security system based on ISO/IEC 27001.

This research shows the analysis of multiple factors that inhibit the implementation of an Information Security Management System (ISMS). The research data were collected from 143 respondents from two universities in northeastern Mexico, in faculties of engineering in related areas. In this study, the Information Security Management System Measurement Instrument (IM-ISMS) was validated. A scale of 24 items was obtained, divided into four factors: organizational policies and regulations, privacy, integrity and authenticity.

This version of the instrument meets the criteria established for its validity (KMO, Bartlett's test of sphericity). An extraction was performed by the minimum residuals method, an oblique rotation was performed by the promax method, when performing the rotation 17 of the 24 items were grouped in the corresponding factor. The final reliability of the scale was calculated by the Omega coefficient, in all dimensions the coefficients were greater than .70, therefore the reliability of the instrument is good.

The results of this study agree with the results found by [10] in which they present a model that complies with ISO/IEC 27002:2013 controls and security and privacy criteria to improve the ISMS. [48], Mentioned that the implementation of controls based on ISO standards can meet the requirements for cybersecurity best practices.

[27], note that models based on ISO 27002 standards allow to diagnose maturity levels in relevant security processes in an organization or to determine what process may be needed and not in practice.

Also, proposing a model with maturity in security indicators can help the cybersecurity auditor to make recommendations to raise the level of security and thus avoid security breaches as pointed out by [49]

The implications of this research are to create and validate an instrument that measures the degree of management of an information security system based on ISO/IEC 27001. Another implication of generating this instrument is to be able to make a diagnosis of the degree of management of an information security system in educational institutions.

## 8    Conclusion

A scale of 24 items was obtained, divided into four factors: organizational policies and regulations, privacy, integrity and authenticity.

This version of the instrument meets the criteria established for its validity (KMO, Bartlett's test of sphericity). An extraction was performed by the minimum residuals method, an oblique rotation was performed by the promax method, when performing the rotation 17 of the 24 items were grouped in the corresponding factor. The final reliability of the scale was calculated by the Omega coefficient, in all the dimensions the coefficients were greater than .70, therefore the reliability of the instrument is good.

**Acknowledgments** The authors wish to thank Damaris Tarango Alvidrez, Vriza Valeria Vazquez Ontiveros and Alejandro García Mendoza, Writing - review & editing.

## References

1.  Olteanu, A. M., Huguenin, K., Shokri, R., Humbert, M., & Hubaux, J. P. (2017). Quantifying interdependent privacy risks with location data. *IEEE Transactions on Mobile Computing*, *16*(3). https://doi.org/10.1109/TMC.2016.2561281
2.  Rincón Soto, I., García Castillo, R. E., & Marín Perea, G. j. (2020). The power of knowledge and information as a generator of value in organizations. *Revista Académica de Investigación*, *11*, 132–147.
3.  Weidman, J., & Grossklags, J. (2019). Assessing the current state of information security policies in academic organizations. *Information & Computer Security*, *28*(3). https://doi.org/10.1108/ICS-12-2018-0142
4.  Valencia-Duque, F. J., & Orozco-Alzate, M. (2017). Metodología para la implementación de un Sistema de Gestión de Seguridad de la Información basado en la familia de normas ISO/IEC 27000. *RISTI - Revista Ibérica de Sistemas e Tecnologias de Informação*, (22). https://doi.org/10.17013/risti.22.73-88
5.  Yoseviano, H. F., & Retnowardhani, A. (2018). The use of ISO/IEC 27001: 2009 to analyze the risk and security of information system assets: case study in xyz, ltd. In *2018 International Conference on Information Management and Technology (ICIMTech)*. IEEE. https://doi.org/10.1109/ICIMTech.2018.8528096
6.  Hoffmann, R., Kiedrowicz, M., & Stanik, J. (2016). Risk management system as the basic paradigm of the information security management system in an organization. *MATEC Web of Conferences*, *76*.
https://doi.org/10.1051/matecconf/20167604010

7. Ahmad, A., Maynard, S. B., Desouza, K. C., Kotsias, J., Whitty, M. T., & Baskerville, R. L. (2021). How can organizations develop situation aware­ness for incident response: A case study of management practice. *Computers & Security*, *101*. https://doi.org/10.1016/j.cose.2020.102122

8. Rehman, H., Masood, A., & Cheema, A. R. (2013). Information Security Management in academic institutes of Pakistan. In *2013 2nd National Con­ference on Information Assurance (NCIA)*. IEEE. https://doi.org/10.1109/NCIA.2013.6725323

9. Benavides Sepúlveda, A., & Blandón Jaramillo, C. (2018). Model infor­mation security management system for entry-level educational institutions. *Scientia Et Technica*, *23*(1), 85–92.

10. Gutiérrez-Martínez, J., Núñez-Gaona, M. A., & Aguirre-Meneses, H. (2015). Business Model for the Security of a Large-Scale PACS, Compliance with ISO/27002:2013 Standard. *Journal of Digital Imaging*, *28*(4). https://doi.org/10.1007/s10278-014-9746-4

11. Solarte Solarte, F. N., Enriquez Rosero, E. R., & Benavides, M. del C. (2015). Metodología de análisis y evaluación de riesgos aplicada a la seguri­dad informática y de la información según la norma ISO/IEC 27001. *Revista Tecnológica ESPOL – RTE*, *28*(5), 492–507. Retrieved from http://www.rte.espol.edu.ec/index.php/tecnologica/article/view/456

12. Miranda Cairo, M., Valdés Puga, O., Pérez Mallea, I., Portelles Cobas, R., & Sánchez Zequeira, R. (2016). Methodology for the implementation of auto­mated management of computer security controls. *Revista Cubana de Cien­cias Informáticas*, *10*(2), 14–26. Retrieved from http://scielo.sld.cu/sci­elo.php?script=sci_abstract&pid=S2227-18992016000200002&lng=es&nrm=iso&tlng=en

13. Jufri, M. T., Hendayun, M., & Suharto, T. (2017). Risk-assessment based ac­ademic information System security policy using octave Allegro and ISO 27002. In *2017 Second International Conference on Informatics and Compu­ting (ICIC)*. IEEE. https://doi.org/10.1109/IAC.2017.8280541

14. Viecco, L. R., & Arevalo, J. G. (2020). Information Technology Governance Model, Based on Risk Management and Information Security for Colombian Public Universities: Case on Study University of La Guajira. *IOP Confer­ence Series: Materials Science and Engineering*, *844*. https://doi.org/10.1088/1757-899X/844/1/012045

15. Khajouei, H., Kazemi, M., & Moosavirad, S. H. (2017). Ranking information security controls by using fuzzy analytic hierarchy process. *Information Sys­tems and e-Business Management*, *15*(1). https://doi.org/10.1007/s10257-016-0306-y

16. Mora, F., Cristina, D., & Guerrero Santander, C. D. (2013). Sistema de ad­ministración de controles de seguridad informática basado en ISO/IEC

27002. *unab.edu.co*. Retrieved June 3, 2021, from https://reposi-tory.unab.edu.co/handle/20.500.12749/3473

17. Pietre-Cambacedes, L., Quinn, E. L., & Hardin, L. (2013). Cyber Security of Nuclear Instrumentation & Control Systems: Overview of the IEC Standardization Activities. *IFAC Proceedings Volumes*, *46*(9). https://doi.org/10.3182/20130619-3-RU-3018.00392

18. Disterer, G. (2013). ISO/IEC 27000, 27001 and 27002 for Information Secu-rity Management. *Journal of Information Security*, *04*(02). https://doi.org/10.4236/jis.2013.42011

19. Alcaraz, C., & Zeadally, S. (2015). Critical infrastructure protection: Re-quirements and challenges for the 21st century. *International Journal of Crit-ical Infrastructure Protection*, *8*. https://doi.org/10.1016/j.ijcip.2014.12.002

20. Montaño Orrego, V. (2011). La gestión en la seguridad de la información según Cobit, Itil e Iso 27000. *Revista Pensamiento Americano*, *2*(6), 21–23. Retrieved from https://dsi.face.ubiobio.cl/sbravo/1-AUDINF/GESTION%20_SEGINF%20.pdf

21. Montaño Ardila, V. M. (2010). Beneficios para el gobierno empresarial: Ar-ticulando COBIT con ISO 27000 para la exitosa implantación de un gobierno de TI. *Económicas CUC*, *31*(31). Retrieved from https://reposito-rio.cuc.edu.co/handle/11323/2900

22. Khanna, P., Zavarsky, P., & Lindskog, D. (2016). Experimental Analysis of Tools Used for Doxing and Proposed New Transforms to Help Organizations Protect against Doxing Attacks. *Procedia Computer Science*, *94*. https://doi.org/10.1016/j.procs.2016.08.071

23. Breier, J., & Hudec, L. (2012). New approach in information system security evaluation. In *2012 IEEE First AESS European Conference on Satellite Tele-communications (ESTEL)*. IEEE. https://doi.org/10.1109/ESTEL.2012.6400145

24. Sánchez Crespo, L. E. (2005). La gestión de la seguridad de los sistemas de información: pasado, presente y futuro. Retrieved June 3, 2021, from https://www.researchgate.net/publication/232252325_SSE-PYME_Desarrollando_herramientas_de_gestion_de_seguridad_para_la_PYME

25. Muyón, C., Guarda, T., Vargas, G., & Ninahualpa Quiña, G. (2019). Es-quema Gubernamental de Seguridad de la Información EGSI y su aplicación en las entidades públicas del Ecuador. *Revista Ibérica de Sistemas e Tecnolo-gias de Informação*, *18*, 310–317. Retrieved from https://www.proquest.com/open-view/f4b193b46ccb16a251428b15a52d084a/1?pq-origsite=gscholar&cbl=1006393

26. Caiza-Acero, M., & Bolaños-Burgos, F. (2014). The implementations of in-formation security standards: a study of case the Sociedad de Lucha Contra el Cáncer del Ecuador. *ReCIBE*, *3*(3).

27. Kurniawan, E., & Riadi, I. (2018). Security level analysis of academic information systems based on standard ISO 27002:2003 using SSE-CMM. *International Journal of Computer Science and Information Security*, *16*(1). Retrieved from https://www.researchgate.net/publication/323029044_Security_level_analysis_of_academic_information_systems_based_on_standard_ISO_270022003_using_SSE-CMM

28. Reyes López, F., Betancurt Domínguez, Y., Muñoz Periñán, I. L., & Paz Loboguerrero, A. F. (2015). Support tool for verifying the compliance of standards and regulations in implementations of strategies for information security. *Sistemas y Telemática*, *13*(32). https://doi.org/10.18046/syt.v13i32.2032

29. Meng, M., & Liu, E. (2015). The Application Research of Information Security Risk Assessment Model Based on AHP Method. *Journal of Advances in Information Technology*. https://doi.org/10.12720/jait.6.4.201-206

30. Kawasaki, R., & Hiromatsu, T. (2014). Proposal of a Model Supporting Decision-Making on Information Security Risk Treatment . *International Scholarly and Scientific Research & Innovation*, *8*(4), 583–589. https://doi.org/http://doi.org/10.5281/zenodo.1092042

31. García Porras, J. C., Huamani Pastor, S. C., & Lomparte Alvarado, R. F. (2018). Modelo de gestión de riesgos de seguridad de la información para PYMES peruanas. *Revista peruana de computación y sistemas*, *1*(1). https://doi.org/10.15381/rpcs.v1i1.14856

32. Imbaquingo, D., Herrera, E., Herrera, I., Arciniega, S. R., Guamán, V. L., & Ortega Bustamante, M. (2019). Evaluación de sistemas de seguridad informáticos universitarios Caso de Estudio: Sistema de Evaluación Docente. *Revista Iberica de Sistemas e Tecnologias de Informacao*, *22*, 349–362. Retrieved from https://www.researchgate.net/publication/338050855_Evaluacion_de_sistemas_de_seguridad_informaticos_universitarios_Caso_de_Estudio_Sistema_de_Evaluacion_Docente

33. Fabrigar, L. R., Wegener, D. T., MacCallum, R. C., & Strahan, E. J. (1999). Evaluating the use of exploratory factor analysis in psychological research. *Psychological Methods*, *4*(3). https://doi.org/10.1037/1082-989X.4.3.272

34. Gravetter, F. J., & Wallnau, L. B. (2013). *Essentials of Statistics for the Behavioral Sciences* (8a ed.). Cengage Learning.

35. Edwards, J. R. (2011). The Fallacy of Formative Measurement. *Organizational Research Methods*, *14*(2). https://doi.org/10.1177/1094428110378369

36. Borsboom, D., Mellenbergh, G. J., & van Heerden, J. (2003). The theoretical status of latent variables. *Psychological Review*, *110*(2). https://doi.org/10.1037/0033-295X.110.2.203

37. Jöreskog, K. G. (1977). Factor analysis by least squares and maximum likelihood methods. *Statistical Methods for Digital Computers*.

38. Lloret-Segura, S., Ferreres-Traver, A., Hernández-Baeza, A., & Tomás-Marco, I. (2014). El análisis factorial exploratorio de los ítems: una guía práctica, revisada y actualizada. *Anales de Psicología*, *30*(3). https://doi.org/10.6018/analesps.30.3.199361

39. Finch, H. (2006). Comparison of the Performance of Varimax and Promax Rotations: Factor Structure Recovery for Dichotomous Items. *Journal of Educational Measurement*, *43*(1). https://doi.org/10.1111/j.1745-3984.2006.00003.x

40. Henson, R. K., & Roberts, J. K. (2006). Use of Exploratory Factor Analysis in Published Research. *Educational and Psychological Measurement*, *66*(3). https://doi.org/10.1177/0013164405282485

41. Matsunaga, M. (2010). How to factor-analyze your data right: do's, don'ts, and how-to's. *International Journal of Psychological Research*, *3*(1). https://doi.org/10.21500/20112084.854

42. Park, H. S., Dailey, R., & Lemus, D. (2002). The Use of Exploratory Factor Analysis and Principal Components Analysis in Communication Research. *Human Communication Research*, *28*(4). https://doi.org/10.1111/j.1468-2958.2002.tb00824.x

43. Preacher, K. J., & MacCallum, R. C. (2003). Repairing Tom Swift's Electric Factor Analysis Machine. *Understanding Statistics*, *2*(1). https://doi.org/10.1207/S15328031US0201_02

44. McDonald, R. P. (1999). *Test theory: A unified treatment*. Lawrence erlbaum associates publishers .

45. Gadermann, A. M., Guhn, M., & Zumbo, B. D. (2012). Estimating ordinal reliability for Likert-type and ordinal item response data: A conceptual, empirical and practical guide. *Practical Assessment, Research and Evaluation*, *17*(3). Retrieved from https://www.researchgate.net/publication/236605201_Estimating_ordinal_reliability_for_Likert-type_and_ordinal_item_response_data_A_conceptual_empirical_and_practical_guide

46. Anderson, J. C., & Gerbing, D. W. (1988). Structural equation modeling in practice: A review and recommended two-step approach. *Psychological Bulletin*, *103*(3). https://doi.org/10.1037/0033-2909.103.3.411

47. Viladrich, C., Angulo-Brunet, A., & Doval, E. (2017). Un viaje alrededor de alfa y omega para estimar la fiabilidad de consistencia interna. *Anales de Psicología*, *33*(3). https://doi.org/10.6018/analesps.33.3.268401

48. Leszczyna, R. (2019). Standards with cybersecurity controls for smart grid-A systematic analysis. *International Journal of Communication Systems*, *32*(6). https://doi.org/10.1002/dac.3910

49. Al-Matari, O. M. M., Helal, I. M. A., Mazen, S. A., & Elhennawy, S. (2020). Adopting security maturity model to the organizations' capability model. *Egyptian Informatics Journal*. https://doi.org/10.1016/j.eij.2020.08.001